\documentclass{jpsj-suppl}
\usepackage{txfonts} 
\usepackage{graphicx,amsmath,wrapfig,epsfig}

\title{Nuclear effects in deep inelastic scattering and transition region}
\author{\textsc{S. Kumano}$^{1,2}$}
\inst{
$^1$KEK Theory Center, Institute of Particle and Nuclear Studies, KEK \\
\ \ 1-1, Ooho, Tsukuba, Ibaraki, 305-0801, Japan \\
$^2$J-PARC Branch, KEK Theory Center,
Institute of Particle and Nuclear Studies, KEK \\
\ \ and Theory Group, Particle and Nuclear Physics Division, 
J-PARC Center \\
\ \ 203-1, Shirakata, Tokai, Ibaraki, 319-1106, Japan
}
\recdate{February 7, 2016}

\abst{
We discuss nuclear effects on neutrino-nuclear interactions 
in a wide kinematical range from shallow to deep inelastic
scattering (DIS) region. There is necessity from neutrino communities
to have precise neutrino-nucleus cross sections 
for future measurements on neutrino oscillations and leptonic CP 
violation. We try to create a model to calculate 
neutrino cross sections in the wide kinematical range, 
from quasi-elastic scattering and resonance productions to the DIS.
In this article, nuclear modifications of structure 
functions are mainly discussed, and a possible 
extension to the $Q^2 \to 0$ region is explained. 
We also comment on the transition region between baryon resonances
and the DIS. There are ongoing experimental efforts on nuclear 
modifications of structure functions or parton distribution functions
such as by pA reactions at RHIC and LHC, Drell-Yan measurements 
at Fermilab, Miner$\nu$a neutrino DIS, and charged-lepton DIS at JLab.
}

\kword{neutrino, nucleon, nucleus, parton, quark, gluon, QCD}

\begin{document}
\maketitle

%%%%%%%%%%%%%%%%%%%%%%%%%%%%%%%%%%%%%%%%%%%%%%%%%%%%%%%%%%%%%%%%%%%%%%%%%%%%%%%%
\section{Introduction}

Neutrino physics is one of major discovery fields in 
nuclear and particle physics as well as astrophysics.
The neutrino oscillation is clearly established,
and efforts are now made for a discovery of leptonic CP violation.
As statistical errors of the neutrino-oscillation measurements
become smaller, it is now important to reduce systematic errors,
which are dominated by neutrino-nucleus interactions.
Of course, nuclear effects could be partially understood by
near-detector experiments; however, it is important to 
know precise neutrino-nucleus interactions theoretically.
This is the purpose of this international workshop on
neutrino-nucleus interactions from MeV to multi-GeV region
toward a better description of the nuclear interactions. 
We theoretically try to provide a code to calculate the neutrino 
cross sections precisely. However, aside from the practical demand 
of neutrino-oscillation experimentalists, neutrinos have been 
playing important roles in hadron and nuclear physics, 
and neutrino-nuclear physics is intrinsically interesting field.
For example, neutrino plays an important role
in supernova explosion, and ultra-high-energy neutrino interactions
are realized in astrophysical circumstances.

\begin{figure}[h!]
\hspace{0.5cm}
\begin{minipage}{0.40\textwidth}
     \includegraphics[width=6.2cm]{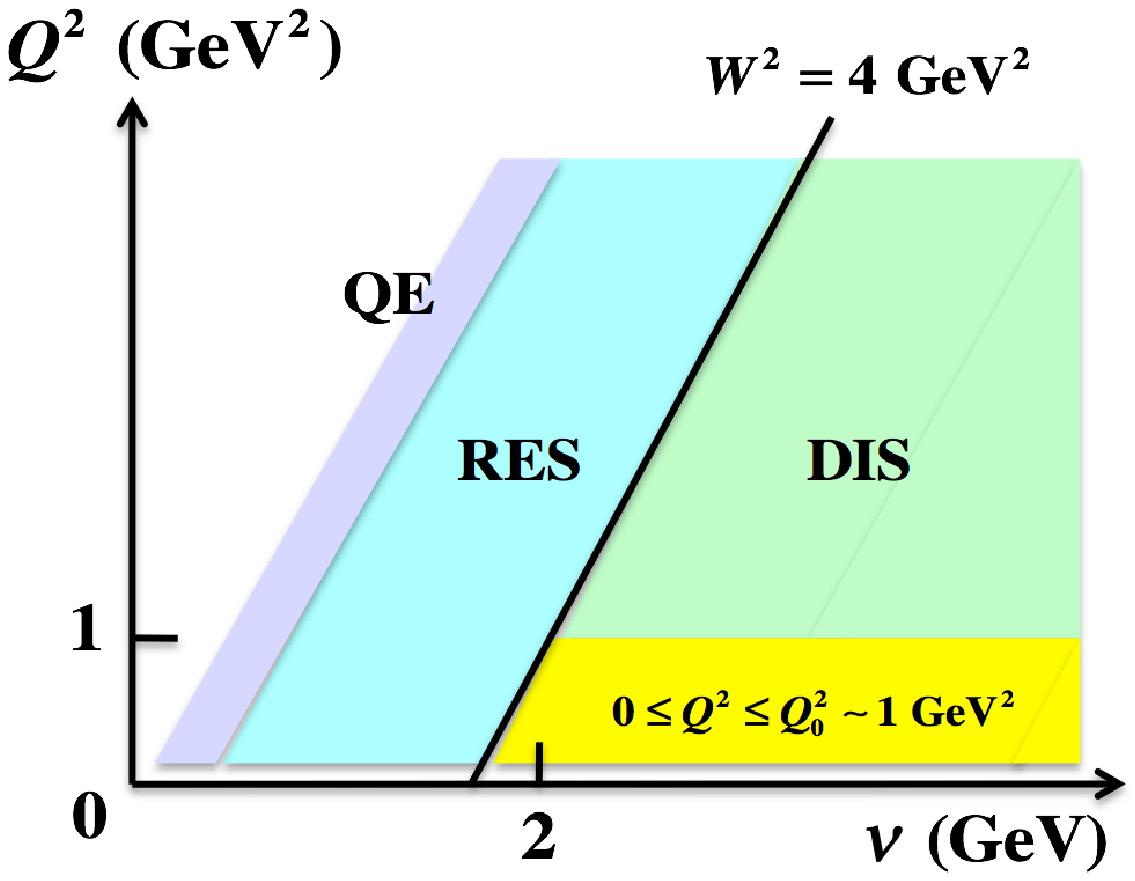}
\vspace{-0.4cm}
\caption{Kinematical regions of neutrino-nucleus scattering.}
\label{fig:qe-res-dis}
\end{minipage}
%%%%%
\hspace{1.0cm}
\begin{minipage}{0.50\textwidth}
   \begin{center}
   \vspace{-0.2cm}
     \includegraphics[width=6.2cm]{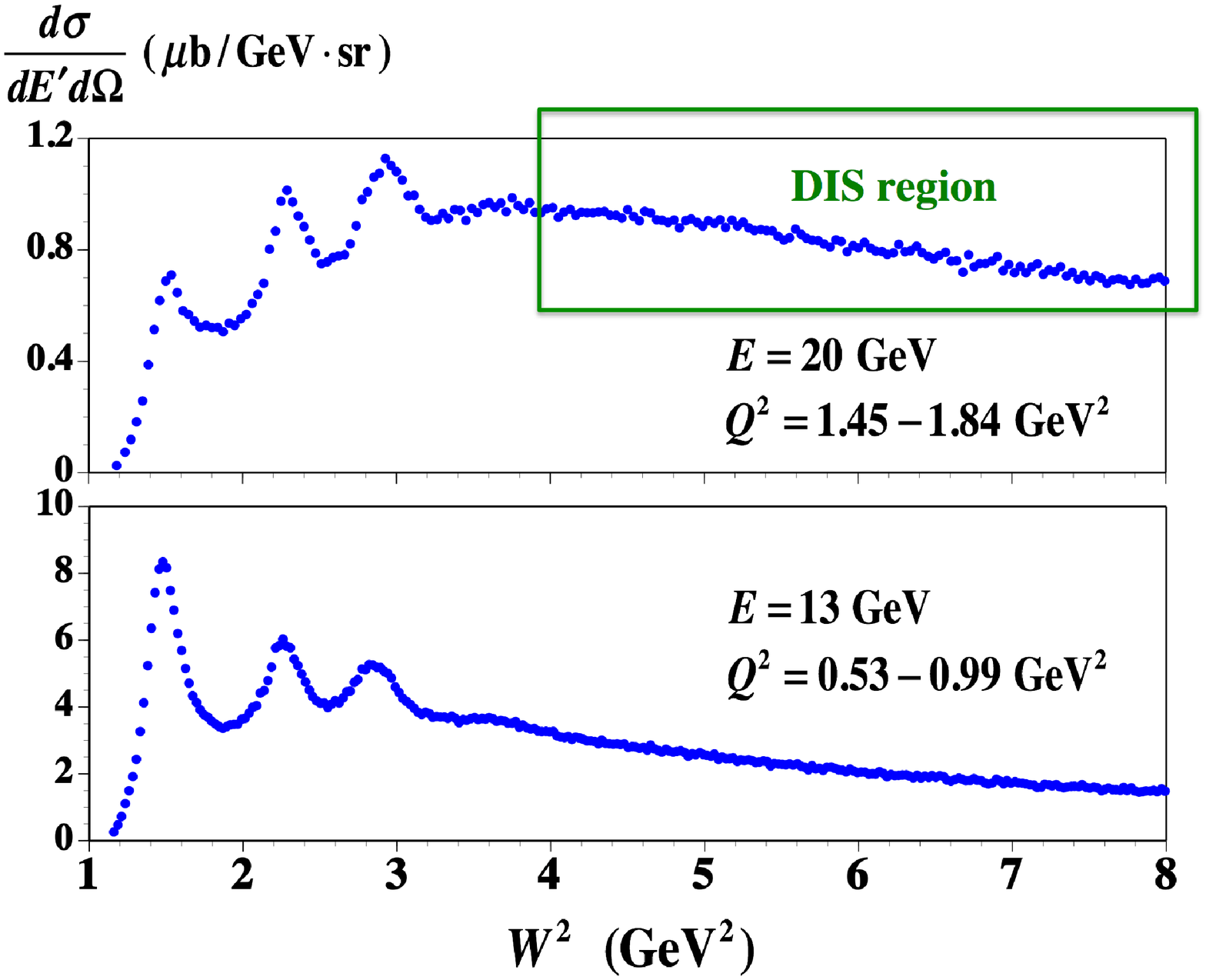}
   \end{center}
\vspace{-0.30cm}
\caption{Electron-proton scattering cross sections \cite{sk-book}.}
\label{fig:res-dis}
\end{minipage} 
\vspace{-1.0cm}
\end{figure}

In this article, we discuss neutrino-nucleus interactions,
especially in the region of deep inelastic scattering (DIS),
for applications to neutrino experiments in the energy region 
from MeV to multi GeV.
There are three major kinematical regions as shown in 
Fig. \ref{fig:qe-res-dis}: quasi-elastic (QE), resonance (RES), 
and DIS. We denote $q$ and $\nu$ as four-momentum and energy
transfers. Then, the elastic scattering occurs if 
$\nu = Q^2 /(2 M_A)$ is satisfied by taking 
$Q^2 = - q^2$ \cite{lepton-kinematics}. At low neutrino energies,
the relation is given by $\nu \simeq \vec q^{\,2} /(2 M_A)$,
which indicates that the energy transfer is equal to
the recoil energy of a nucleus. 
This process is elastic scattering,
and the neutrino sees the nucleus as a whole system.
%%%%%
As the neutrino energy increases, the Compton wavelength
becomes short enough to resolve individual nucleons
in the nucleus, and coherence of many nucleons is no longer
important. Then, the kinematical relation is given by
$\nu = Q^2 /(2 M_N) \simeq \vec q^{\,2} /(2 M_N)$
if the nucleon were at rest. For a rough estimate on
the nucleon kinematics in a nucleus, we may assume that
the binding effect is effectively included into 
a modified nucleon mass $M_N^{\, *}$,
then the relation is
$\nu \simeq (\vec p + \vec q \, )^{2} /(2 M_N^{\, *})
            - \vec p^{\,2} /(2 M_N^{\, *})$
with the nucleon momentum $\vec p$.
It becomes
$\vec q^{\,2} - 2 \, q \, p_{_F} \le 2 M_N^{\, *} \nu
 \le \vec q^{\,2} + 2 \, q \, p_{_F}$ if it is expressed by
the nucleon's Fermi momentum $p_{_F}$.
This is the quasi-elastic region shown in Fig. \ref{fig:qe-res-dis}.
As the energy increases, nucleon resonances start to appear,
and this kinematical region is roughly given by 
the invariant-mass squared as
$M_N^2 < W^2 = (p+q)^2 < 4$ GeV$^2$. This is the
resonance region in Fig. \ref{fig:qe-res-dis}.
In the region $W^2 = (p+q)^2 \ge 4$ GeV$^2$, the nucleon 
is broken up into hadron pieces, and it is 
the deep-inelastic-scattering region. Here, the reaction is
described by introducing partons with incoherent impulse
approximation for scattering processes from individual partons.
For discussing partons with the DGLAP $Q^2$ evolution equations,
the running coupling constant $\alpha_s (Q^2)$ should be
small enough. It requires $Q^2$ should be large,
typically $Q^2 \ge 1$ GeV$^2$ or a few GeV$^2$.
In addition, there is a region which does not fit into any 
of these three categories in Fig. \ref{fig:qe-res-dis}. 
It is the region, $W^2 \ge 4$ GeV$^2$ and $Q^2 < 1$ GeV$^2$.
Here, there is a guideline to describe the axial-vector part
of structure functions by the partial conservation of 
the axial-vector current in the limit $Q^2 \to 0$,
whereas the current is conserved in the vector part.
All these kinematical regions need to be understood precisely
for an application to neutrino reactions, and such an effort 
is in progress for calculating neutrino-nucleus cross sections
in any kinematics as shown in Fig. \ref{fig:qe-res-dis}
\cite{J-PARC-th-neutrino-A}.

Actual cross sections are shown in Fig. \ref{fig:res-dis}
for electron-proton scattering as the function of $W^2$
\cite{sk-book}.
Resonances bumps are clear in the small $W^2$ region, and
the cross sections are continuous at $W^2 \ge 4$ GeV$^2$,
which is the DIS region. If the cross sections are
plotted as a function of $Q^2$ at fixed $W^2$, they are almost
constant. This fact suggests the existence of point-like partons.
%%%%%
In this article, we discuss mainly the DIS region and 
comment on a possible extension to the $Q^2 < 1$ GeV$^2$ region
and also a connection to the resonance region.
Charged-lepton DIS measurements have been done extensively
for the proton target, and its structure functions $F_2$ and
$F_L$ (or $R=F_L/F_T$) are determined experimentally.
Using the parton picture together with the DGLAP evolution
and higher-order $\alpha_s$ corrections, we can extract
parton distribution functions (PDFs) from a global analysis
of the $F_2$ data and other high-energy reaction data of
the proton. Then, the PDFs are used for predicting nucleon's
structure functions in the neutrino DIS and subsequently
neutrino-nucleon cross sections.

In neutrino oscillation experiments, for example, in T2K,
the target is water which contains the oxygen nucleus.
Therefore, nuclear corrections are properly calculated
for precise neutrino-nucleus (oxygen) cross sections.
If we consider a nucleus as a collections of partons
in the same way with the nucleon, this precise cross-section
calculation means that nuclear parton distribution functions (NPDFs)
need to be determined instead of the well-known nucleonic PDFs.
In the following, we discuss neutrino-nucleus DIS 
and NPDF determination in Sec.\,\ref{dis}.

\vspace{-0.60cm}

%%%%%%%%%%%%%%%%%%%%%%%%%%%%% figure %%%%%%%%%%%%%%%%%%%%%%%%%%%%%
%%%%%%%%%%%%%%%%%%%%%%%%%%%%% figure %%%%%%%%%%%%%%%%%%%%%%%%%%%%%

%%%%%%%%%%%%%%%%%%%%%%%%%%%%%%%%%%%%%%%%%%%%%%%%%%%%%%%%%%%%%%%%%%%%%%%%%%%%%%%%
\section{Neutrino-nucleus scattering in deep inelastic region}
\label{dis}
\vspace{-0.2cm}

We explain first the cross section and structure functions
in neutrino-nucleus (or nucleon) scattering in the DIS kinematics
in Sec.\,\ref{sf}.
The structure functions are expressed in terms of the PDFs.
Then, nuclear modifications of the PDFs are discussed in Sec.\,\ref{npdfs}.

%%%%%%%%%%%%%%%%%%%%%%%%%%%%%%%%%%%%%%%%%%%%%%%%%%%%%%%%%%%%%%%%%%%%%%%%%%%%%%%%
\subsection{Structure functions in neutrino-nucleus scattering}
\label{sf}

The neutrino-nucleon (or nucleus) cross section is expressed
in terms of three structure functions, $F_1$, $F_2$, and $F_3$, 
for a charged-current (CC) interaction as
\begin{align}
\frac{{d\sigma _{\nu ,\bar \nu }^{CC}}}{{dx{\rm{ }}dy}}
= 
\frac{{G_F^2{\rm{ }}ME}}{{\pi}}
\left[ \, {x \, {y^2}F_1^{CC} + 
\left( {1 - y - \frac{{M{\rm{ }}x\,y}}{{2 E}}} \right)
{\rm{ }}F_2^{CC} \pm x\,y
\left( {1 - \frac{y}{2}} \right){\rm{ }}F_3^{CC}} \, \right] ,
\end{align}
where $M$ is the mass of the nucleon or a nucleus, 
$E$ is the initial neutrino energy, $G_F$ is the Fermi coupling constant,
$y$ is defined by $y = p \cdot q / p \cdot k = \nu /E$
with the initial neutrino momentum $k$,
and $\pm$ indicates $+$ for $\nu$ and $-$ for $\bar \nu$.
The structure function $F_3$ does not exist in the charged-lepton DIS,
and it has new information especially for constraining valence-quark 
distributions $q_v = q - \bar q$. The minus sign in front of $\bar q$
stems from the fact that the antiquark parity is opposite to 
the quark one.

In the DIS region, the virtual $W$ or $Z$ interacts with a parton
in a short time, which is much shorter than parton interaction time
estimated by the uncertainty principle, the cross section is described
by the incoherent sum of individual parton cross sections.
Their expressions are given in the leading order (LO) of $\alpha_s$ as
\begin{align}
2xF_1 (x,Q^2)_{\rm LO} & = F_2 (x,Q^2)_{\rm LO}, 
\nonumber \\
  F_2^{\,\text{CC}(\nu p)}(x,Q^2)_{\rm LO} & = 2 x \left [ \, d(x,Q^2) + s(x,Q^2) 
                           + \bar u (x,Q^2) + \bar c (x,Q^2) \, \right ], 
\label{eqn:f123-lo}
\\
x F_3^{\,\text{CC}(\nu p)}(x,Q^2)_{\rm LO} & = 2 x \left [ \, d(x,Q^2) + s(x,Q^2) 
                           - \bar u (x,Q^2) - \bar c (x,Q^2) \, \right ] .
\nonumber
\end{align}
In actual calculations of the structure functions and then the cross sections,
higher-order $\alpha_s$ corrections are included in the form of the coefficient
functions, $C_n^{\,q}$ and $C_n^{\,g}$, as \cite{qcd-collider-book}
\begin{align}
\bar F_n (x,Q^2)  =  C_n^{\,q} (x,Q^2) \otimes \bar F_n (x,Q^2)_{\rm LO}
                       + C_n^{\,g} (x,Q^2) \otimes xg(x,Q^2), 
\ \ \ n = 1,\ 2, \ 3 ,
\label{eqn:f123}
\end{align}
where $\bar F_n$ is given by
$\bar F_1  = x F_1$, $\bar F_2 = F_2$, and $\bar F_3 = x F_3$,
and $\otimes$ indicates the convolution integral
$f (x) \otimes g (x) = \int _x^1 (dy / y) f (x/y) g(y)$.
In the $Q^2$ region of a few GeV$^2$, there are significant
higher-twist contributions to the structure functions. Therefore,
$F_1$ is calculated from $F_2$ by the relation
\begin{equation}
2 x F_1(x,Q^2) = \frac{1 + 4M^2 x^2/Q^2}{1 + R(x,Q^2)} F_2(x,Q^2) ,
\label{eqn:f1-r-f2}
\end{equation}
with the longitudinal-transverse structure function ratio 
$R = F_L / (2xF_1)$.
Experimental measurements are usually shown by $F_2$ and $F_3$
together with Eq.\,(\ref{eqn:f1-r-f2}) by analyzing cross-section data.
Then, using the $F_2$ and $F_3$ data and other ones,
one obtains the optimum PDFs. Therefore, although the expression 
Eq.\,(\ref{eqn:f123}) is theoretically consistent for $F_1$, 
there could be some differences from experimental measurements at small $Q^2$.
If we supply information on the PDFs of the nucleon or a nucleus,
the structure functions and the cross sections are calculated.
The nucleonic PDFs are determined by a global analysis of high-energy 
scattering data on the proton. Because of abundant measurements
in charged-lepton DIS, neutrino DIS, Drell-Yan, and so on,
the nucleonic PDFs are now well determined except for extreme kinematical
regions, which is a base for a new discovery in LHC experiments.
For the neutrino interactions, we have an additional complication
from nuclear medium effects on the PDFs. This topic is discussed
in the next section.

%%%%%%%%%%%%%%%%%%%%%%%%%%%%%%%%%%%%%%%%%%%%%%%%%%%%%%%%%%%%%%%%%%%%%%%%%%%%%%%%
\subsection{Nuclear parton distribution functions}
\label{npdfs}

Nuclear modifications of the PDFs can be obtained by analyzing
measurements on nuclear structure function ratios such as
$F_2^A /F_2^D$, where $A$ and $D$ indicate a nucleus and 
the deuteron, respectively. The ratios $F_2^A /F_2^D$ are measured
in charged-lepton DIS from small nuclei to large ones. Furthermore,
there are available measurements on the Drell-Yan cross ratios
$\sigma_{pA} / \sigma_{pD}$, which could constrain antiquark
nuclear modifications. We also obtain information from high-energy
nuclear reactions at RHIC and LHC.

%%%%%%%%%%%%%%%%%%%%%%%%%%%%%%%%%%%%%%%%%%%%%%%%%%%%%%%%%%%%%%%
\begin{wrapfigure}[13]{r}{0.42\textwidth}
   \vspace{-0.80cm}
\begin{center}
   \includegraphics[width=0.40\textwidth]{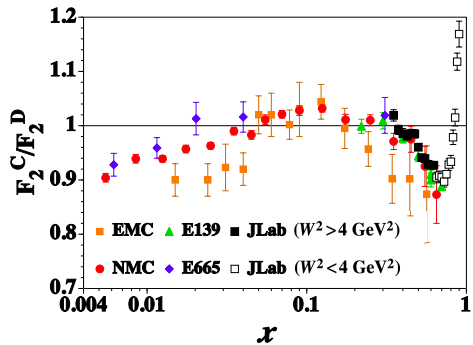}
\end{center}
    \vspace{-0.35cm}
\caption{Charged-lepton DIS measurements for $F_2^C /F_2^D$.}
\label{fig:emc-f2}
\end{wrapfigure}
%%%%%%%%%%%%%%%%%%%%%%%%%%%%%%%%%%%%%%%%%%%%%%%%%%%%%%%%%%%%%%%

As an example, measurements are shown in Fig. \ref{fig:emc-f2}
for nuclear modifications of $F_2$ structure functions
$F_2^C /F_2^D$ in charged-lepton DIS. 
The large-$x$ ($>0.65$) JLab data are shown by the open boxes
because their invariant-mass squared are smaller than 4 GeV$^2$.
The nuclear modifications are positive or negative depending
on the $x$ region. At large $x$, the ratio increases as $x \to 1$,
although the data are not in the DIS region, and the increase
is explained by the nucleon Fermi-motion in the nucleus.
At medium $x$ ($0.3 < x <0.7$), the ratio is smaller than one.
It is caused by nuclear binding and possible nucleon 
modification in a nuclear medium. At small $x$ ($<0.05$),
the effects are negative due to nuclear shadowing. In this region,
the virtual photon propagates as a $q \bar q$ pair 
or a vector meson to a significant distance scale more than
the average nucleon separation 2.2 fm in a nucleus, 
and this reaction is described by a multiple scattering theory.
The $q\bar q$ or vector meson interacts 
strongly in the surface region of the nucleus, and
the reaction causes nuclear shadowing. Namely, internal nucleons 
are shadowed due to the interactions with the surface ones.
In the intermediate region ($0.1<x<0.2$), the modifications are
positive, and it is called anti-shadowing. The NPDFs should 
satisfy the conservations of charge, baryon number, and momentum,
so that there should be positive effects if there are negative
contributions at small $x$ and medium $x$. Contributions
from the large-$x$ ($>0.9$) region are very small to these
conservation conditions because the NPDFs themselves are tiny.
For the theoretical mechanisms of these descriptions, one may
read summary papers, for example, in Ref. \cite{nuclear-summary}.

We can extract the NPDFs from the measurements of the $F_2^A$ ratios,
Drell-Yan ($p + A \to \mu^+ \mu^- +X$), and other data from $pA$
and $AA$ reactions at RHIC and LHC. First, the NPDFs are expressed
by a number of parameters which are then determined by a $\chi^2$
analysis of world data on high-energy nuclear reactions.
It is easier to obtain the nuclear modifications from the nucleonic
PDFs because measurements are often shown by the form of
nuclear ratios as given in Fig. \ref{fig:emc-f2}. However,
a parametrization of the NPDFs themselves is also possible,
as an alternative method, in the same way with 
the nucleonic PDF determination.
In any case, both results should agree with each other in principle.
For example, the NPDFs could be expressed as
\begin{equation}
f_i^A (x,Q_0^2) = w_i (x,A,Z) \, 
    \frac{1}{A} \left[ Z\,f_i^{p} (x,Q_0^2) 
                + N f_i^{n} (x,Q_0^2) \right] ,
\label{eqn:fia}
\end{equation}
where $p$ and $n$ indicate proton and neutron, 
$A$, $Z$, and $N$ are mass number, atomic number, and neutron number,
and $i$ indicates the parton type ($i=u_v,\ d_v,\ \bar q,\ g$).
If there were no nuclear modification, the NPDFs are given by
a simple summation of proton and neutron contributions.
It corresponds to $w_i (x,A,Z)=1$ in Eq.\,(\ref{eqn:fia}).
The scale $Q_0^2$ is a fixed $Q^2$ point to define the NPDFs, 
and the $Q^2$ evolution from $Q_0^2$
is evaluated by the standard DGLAP evolution equations.
The kinematical region of the scaling variable is $0<x<A$ for
the nucleus with the mass number $A$. The expression of 
Eq.\,(\ref{eqn:fia}) and also the DGLAP evolution equations
cannot handle the $x>1$ region. However, we do not consider
this region for the time being because there is no DIS data 
in such a region and the distributions are extremely small
in any case.

%%%%%%%%%%%%%%%%%%%%%%%%%%%%%%%%%%%%%%%%%%%%%%%%%%%%%%%%%%%%%%%
\begin{wrapfigure}[13]{r}{0.42\textwidth}
   \vspace{+0.30cm}
   \hspace{-0.10cm}
   \includegraphics[width=0.41\textwidth]{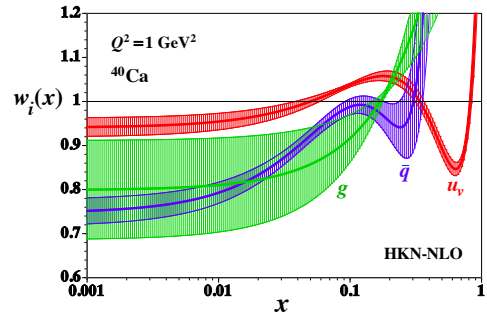}
    \vspace{-0.0cm}
\caption{Typical nuclear modifications of PDFs for $^{40}$Ca
         at $Q^2=1$ GeV$^2$.}
\label{fig:wx}
\end{wrapfigure}
%%%%%%%%%%%%%%%%%%%%%%%%%%%%%%%%%%%%%%%%%%%%%%%%%%%%%%%%%%%%%%%

Typical nuclear modifications are shown in Fig.\,\ref{fig:wx}
for the $^{40}$Ca nucleus at $Q^2=1$ GeV$^2$ with uncertainty
bands \cite{some-npdfs}. Because the mass number $A=40$ is larger 
for the calcium than $A=12$ for the carbon, the nuclear modifications
are generally larger than the ones in Fig.\,\ref{fig:emc-f2}.
In the region $x>0.3$, the structure function $F_2$ is dominated
by the valence-quark distributions, so that the modifications
of $u_v (x)$ resemble to the ones for $F_2$ at $x>0.3$
in Fig.\,\ref{fig:emc-f2}. In the same way, $F_2$ is dominated
by the antiquark distributions at small $x$, so that the modifications
of $\bar q$ are controlled by the shadowing data of $F_2$ at $x<0.05$.
The antiquark distributions in the region of $0.1<x<0.2$ are
fixed by Fermilab Drell-Yan measurements, so that the anti-shadowing
of $F_2$ is given by the positive modification of the valence-quark
distributions at $0.1<x<0.2$. There is no strong constraint for
the gluon shadowing at small $x$ and also for gluon distributions
at large $x$ at this stage. There are a few groups on the analysis 
of NPDFs, but obtained NPDFs are similar \cite{some-npdfs}. 
The only noticeable differences exist in the gluon modifications
and antiquark ones at large $x$ ($>0.3$).

%%%%%%%%%%%%%%%%%%%%%%%%%%%%%%%%%%%%%%%%%%%%%%%%%%%%%%%%%%%%%%%
\begin{wrapfigure}[12]{r}{0.42\textwidth}
   \vspace{-0.70cm}
   \hspace{-0.15cm}
   \includegraphics[width=0.41\textwidth]{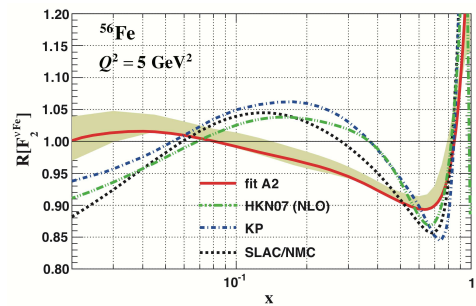}
   \vspace{-0.2cm}
\caption{Nuclear modifications of $F_2^{\nu \text{Fe}}$\\
in comparison with charged-lepton ones \cite{nCTEQ-nu}.}
\label{fig:neutrino-charged-lepton}
\end{wrapfigure}
%%%%%%%%%%%%%%%%%%%%%%%%%%%%%%%%%%%%%%%%%%%%%%%%%%%%%%%%%%%%%%%

Using determined NPDFs, one can calculate nuclear modifications
of the structure functions in the neutrino DIS. The NPDF combinations
in Eqs. (\ref{eqn:f123-lo}) and (\ref{eqn:f123}) are different 
from the ones of charged-lepton DIS, so that the nuclear 
modifications of $F_2$ and $F_3$ are slightly different
from the ones in Fig.\,\ref{fig:emc-f2}. However, we do not 
expect a large difference for $F_2$ from the modifications 
of the charged-lepton DIS in Fig.\,\ref{fig:emc-f2}.
In spite of this expectation, noticeable differences are reported
by the nCTEQ collaboration in Ref.\,\cite{nCTEQ-nu} as shown
in Fig.\,\ref{fig:neutrino-charged-lepton}.
The solid curve with uncertainties indicates nuclear modifications
suggested by analyzing neutrino DIS data, whereas the other three curves
are modifications from analysis of charged-lepton DIS data.
There are significant differences: 
(1) the medium-$x$ ($\sim 0.6$) depletion is smaller,
(2) there is no anti-shadowing at $x \sim 0.2$,
(3) there is no shadowing indication at small $x \ (<0.05)$.
However, there is no such obvious difference in the analysis of
Eskola {\it et al.} and de Florian {\it et al.} \cite{some-npdfs},
so that detailed independent studies are needed to clarify
the situation. In any case, the Miner$\nu$a experiment is going on
at Fermilab to investigate the nuclear modifications in the neutrino DIS,
so that clear information could be obtained for this issue from 
the experimental side.
On the other hand, existing measurements of charged-lepton
and neutrino structure functions are compared directly,
and the results seem to be consistent except for the small-$x$ region
($x<0.05$) \cite{Kalantarians-2015}.

Aside from this issue, the NPDFs are determined by various groups
from global analyses of high-energy nuclear reaction data,
and useful codes are available for calculating the NPDFs at any
kinematics as long as it is in the DIS region \cite{some-npdfs}.
As obvious from Fig.\,\ref{fig:wx}, the gluon distributions 
are not well determined in nuclei. 
At small $x$, the gluon distribution is much larger than
antiquark and valence-quark distributions, so that a nucleus
should be considered as a gluonic many-body system in high-energy
reactions. The lack of our knowledge on the gluon shadowing could
affect understanding on properties of quark-gluon plasma 
and also a possible new discovery in heavy-ion collisions at LHC.
The determined NPDFs are used in Eqs.\,(\ref{eqn:f123-lo}),
(\ref{eqn:f123}), and (\ref{eqn:f1-r-f2})
for calculating the neutrino DIS structure functions and then
the neutrino cross sections. Therefore, a precise determination
of the NPDFs is very important not only for heavy-ion physics
but also for other projects such as neutrino-oscillation 
and leptonic-CP-violation physics.

%%%%%%%%%%%%%%%%%%%%%%%%%%%%%%%%%%%%%%%%%%%%%%%%%%%%%%%%%%%%%%%%%%%%%%%%%%%%%%%%
\section{Neutrino-nucleus scattering in transition regions}
\label{transition}

%%%%%%%%%%%%%%%%%%%%%%%%%%%%%%%%%%%%%%%%%%%%%%%%%%%%%%%%%%%%%%%%%%%%%%%%%%%%%%%%
\subsection{Neutrino-nucleus scattering at $Q^2 \to 0$ and $W^2 \ge 4$ GeV$^2$}
\label{regge}

We briefly explain descriptions of the region,
$0 < Q^2 < 1$ GeV$^2$ and $W^2 \ge 4$ GeV$^2$,
in Fig.\,\ref{fig:qe-res-dis}. This region cannot be described by partons
because of the kinematical condition $Q^2 < 1$ GeV$^2$
and also by nucleon resonances due to the condition $W^2 \ge 4$ GeV$^2$. 
Theoretical descriptions of this regions are explained, for example,
in Ref.\,\cite{Q2->0-theory}.
The transverse and longitudinal structure functions are defined by
corresponding total neutrino cross sections $\sigma_{T,\,L}$ as
\begin{align}
F_{T,\,L} (x,Q^2) & = \frac{1}{\pi} \, \sqrt{1+Q^2/\nu^2}
                    \, Q^2 \, \sigma_{T,\,L} \, ,
\nonumber \\
\sigma_{T,\,L}  & = \frac{(2 \pi)^4}{4\sqrt{(p\cdot q)^2-p^2 q^2}}
     \sum_f  \delta^{\,4} (p+q-p_f) 
     \left | \, \langle \, f \, | \, \varepsilon_{T,\,L} \cdot J (0) \, 
           | \, p \, \rangle \,  \right |^2 ,
\label{eqn:f-tl}
\end{align}
where $\varepsilon_{T,\,L}$ is the polarization vector of $W$ or $Z$.
The transverse cross section is finite even at $Q^2 =0$, which indicates
$F_T \sim Q^2$ from Eq.\,(\ref{eqn:f-tl}), so that it vanishes 
($F_T \to 0$) as $Q^2 \to 0$. The longitudinal part should be
carefully taken care of due to non-conservation of the axial current.
In weak interactions, the current consists of vector and axial-vector
components. The vector current is conserved, so that
the vector-current part of the hadron tensor should satisfy 
$q^{\,\mu} W_{\mu\nu} =0$. Therefore, we have the relation
$F_L^{\,V} \sim Q^2 F_T^{\,V}$ 
in the same way with the charged-lepton DIS.

As for the axial-vector current, it is not conserved and there is 
a constraint by the partial conservation of the axial-vector current
(PCAC), $\partial_\mu \, A_a ^{\,\mu} (x) = f_\pi \, m_\pi^2 \, \pi_a (x)$,
where $f_\pi$ is the pion-decay constant, $m_\pi$ is the pion mass,
and $\pi_a (x)$ is the pion field with the isospin index $a$.
This PCAC leads to the relation that the axial-vector part of the
structure function is finite at $Q^2 \to 0$, and it is 
given by the pion-scattering cross section 
$\sigma_\pi$ as $F_L^{\,A} \to f_\pi^{\,2} \, \sigma_\pi /\pi$ 
in the limit $Q^2 \to 0$.

From these considerations, we provide structure functions at small $Q^2$
by separating them into two parts, vector and axial-vector ones, and
we impose the PCAC for the axial-vector one. Then, the neutrino cross
section should be described in the region, $Q^2 \to 0$ and $W^2 \ge 4$ GeV$^2$,
in addition to the one in the DIS region. A simple parametrization was
proposed in Ref. \cite{by-2011} for connecting the DIS region to the one
in $Q^2 < 1$ GeV$^2$ by using the knowledge of the DIS structure functions
(or the PDFs) and the PCAC. In the simulation code of FLUKA \cite{fluka-2009},
the PCAC is not used and it is simply assumed as 
$F_{2,\, 3} (x,Q^2) = (2 Q^2)/(Q_0^2 + Q^2) \, F_{2,\, 3} (x,Q_0^2)$
at $Q<Q_0^2$.

%%%%%%%%%%%%%%%%%%%%%%%%%%%%%%%%%%%%%%%%%%%%%%%%%%%%%%%%%%%%%%%%%%%%%%%%%%%%%%%%
\subsection{Neutrino-nucleus scattering at $W^2 < 4$ GeV$^2$}
\label{resonace}

The region of $W^2 < 4$ GeV$^2$ is described by baryon resonances,
and this topic is covered by other speakers at this workshop.
If there is a good theoretical model for describing the resonances
up to the invariant-mass squared $W_0^2$ ($\sim 4$ GeV$^2$)
and if the structure functions are described in the DIS region 
at $W^2 \ge W_0^2$, we may simply combine
these two models at the boundary of $W_0^2$.
Then, the combined code can be used for calculating the neutrino-nucleus 
cross sections.

On the other hand, there is an interesting topic of hadron duality
in this transition region \cite{summary-duality}.
At low energies, a cross section is described in terms of hadron 
degrees of freedom, whereas quark and gluon degrees of freedom
should be used at high energies. However, there is an interesting 
phenomenon called Bloom-Gilman duality:
if structure functions in the resonance region are averaged over
a scaling variable, they are almost equal to the DIS ones.
There are also QCD studies on the hadron duality in terms of
the operator product expansion for the moments of structure functions.
There is a revived interest on this topic because JLab produces 
measurements at large $x$ in the transition $Q^2$ and $W^2$ 
region with much better accuracies than
the SLAC measurements in 1970's.

\vfill\eject

%%%%%%%%%%%%%%%%%%%%%%%%%%%%%%%%%%%%%%%%%%%%%%%%%%%%%%%%%%%%%%%
\begin{wrapfigure}[20]{r}{0.48\textwidth}
   \vspace{+0.00cm}
\begin{center}
   \includegraphics[width=0.45\textwidth]{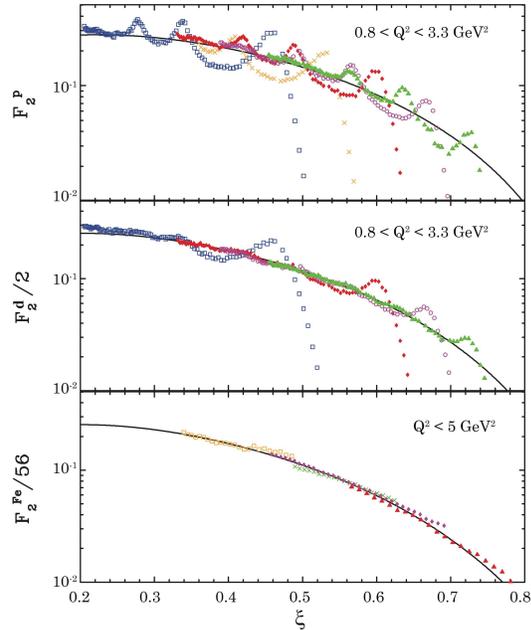}
\end{center}
    \vspace{-0.25cm}
\caption{Structure functions of proton, deuteron, and iron 
\cite{summary-duality}.}
\label{fig:duality-A}
\end{wrapfigure}
%%%%%%%%%%%%%%%%%%%%%%%%%%%%%%%%%%%%%%%%%%%%%%%%%%%%%%%%%%%%%%%

As an example, the structure functions of proton, deuteron, and iron
are shown in Fig.\,\ref{fig:duality-A} as the function of 
the Nachtmann scaling variable $\xi = 2x /(1+\sqrt{1+4M^2x^2/Q^2})$
in comparison with experimental measurements in the resonance region.
The structure function curves are obtained by using the GRV parametrization
in 1998 with nuclear corrections. Actually, the DIS structure functions
are almost equal to the resonance ones averaged over the scaling variable.
The resonance bumps are clear in the proton; however, they are washed
out in the iron nucleus due to the nucleon's Fermi motion.
Along the line of this kind of studies, the Bodek-Yang parametrization
is widely used in neutrino studies \cite{by-2011} by extending the DIS
structure functions to the region of $W^2 < 4$ GeV$^2$.
In this case, it is not possible to describe the details of
the $\xi$ dependence in the neutrino-nucleus (or nucleon) 
cross sections, but the averaged properties are reproduced 
by this model as shown in Fig.\,\ref{fig:duality-A}. 

%%%%%%%%%%%%%%%%%%%%%%%%%%%%%%%%%%%%%%%%%%%%%%%%%%%%%%%%%%%%%%%%%%%%%%%%%%%%%%%%
\section{Experimental progress on nuclear structure functions}
\label{exp-progress}

Since there are measurements on high-energy pA and dA reactions 
at LHC, the NPDFs should be more constrained especially at small $x$ 
by including their data. Particularly, the situation of the gluon 
shadowing should be improved. There are comparisons with 
the LHC data on charged-hadron, dijets, and direct-photon 
productions in p+Pb. At this stage, the measurements are consistent 
with the existing NPDFs \cite{lhc-npdfs}.
In addition to the new information from LHC, an updated analysis
in progress at JLab for their measurements on $F_2^A/F_2^D$, 
and they are useful for constraining the nuclear modifications 
of valence-quark distributions at medium $x$.
The Fermilab Drell-Yan measurements played an important role
in constraining the antiquark modifications at $x \sim 0.1$.
Now, the Fermilab-E906 experiment is in progress for measuring
pA/pd cross-section ratios for the Drell-Yan process in an extended
$x$ region, in addition to the measurement on $\bar u(x)/\bar d(x)$
\cite{fermilab-e906,ubar-dbar}. As found in Fig.\,\ref{fig:wx},
there is no constraint for the antiquark modifications at $x>0.2$, 
and this situation should be improved in the near future.

In neutrino experiments relevant for this workshop,
there is ongoing analysis by the Miner$\nu$a collaboration 
on measurements of carbon, iron, and lead structure functions
in the DIS region \cite{minerva}. So far, observed modifications
on C/CH, Fe/CH, and Pb/CH are mostly consistent with current
simulation results. However, there are suppressions at 
$0.05 < x < 0.2$ in the lead data. Since the range $0.1 < x < 0.2$
is considered as the anti-shadowing region, it is
difficult to understand the suppression at this stage.
In any case, the shadowing mechanism could be different
for the weak interactions in the axial-vector part
because axial-vector mesons (or corresponding $q\bar q$ states)
contribute to the shadowing in a different way.
Although the Miner$\nu$a experiment will not 
probe the region of small enough $x$, its analysis result 
could provide us some hint toward a new shadowing phenomenon
different from the charged-lepton one. In any case, it is 
sensitive to the region $0.05<x<0.6$ including
the anti-shadowing range $0.1 < x < 0.2$.
There are many theory articles at small $x$ ($<0.05$), 
medium $x$ ($0.3<x<0.7$), and large $x$ ($>0.7$),
so that major physics mechanisms are known in these regions.
However, the mechanism on the anti-shadowing is not well investigated,
although such a positive modification is kinematically obvious 
from the conservations of baryon-number, charge, and momentum.
Therefore, the Miner$\nu$a experiment could shed light 
on the mechanism of anti-shadowing.

%%%%%%%%%%%%%%%%%%%%%%%%%%%%%%%%%%%%%%%%%%%%%%%%%%%%%%%%%%%%%%%%%%%%%%%%%%%%%%%%
\section*{Acknowledgements}
\vspace{-0.3cm}
This work was supported by Ministry of Education, Culture, Sports, 
Science and Technology (MEXT) KAKENHI Grant No. 25105010.
The author thanks R. Ent, C. E. Keppel, W. Melnitchouk, and
J. G. Morfin for supplying their figures in this article.

%%%%%%%%%%%%%%%%%%%%%%%%%%%%%%%%%%%%%%%%%%%%%%%%%%%%%%%%%%%%%%%%%%%%%%%%%%%%%%%%

%%%%%%%%%%%%%%%%%%%%%%%%%%%%%%%%%%%%%%%%%%%%%%%%%%%%%%%%%%%%%%%%%%%%%%%%%%%%%%%%


\begin{thebibliography}{9}
\bibitem{lepton-kinematics} T. De Forest and J. D. Walecka, 
           Adv. Phys. {\bf 15} (1966) 1.
\bibitem{J-PARC-th-neutrino-A} See http://nuint.kek.jp/
   for activities of neutrino-nucleon interaction collaboration
   at J-PARC branch of KEK theory center;
   Y. Hayato, M. Hirai, W. Horiuchi, H. Kamano, S. Kumano, 
   T. Murata, S. Nakamura, K. Saito, M. Sakuda, and T. Sato, 
   to be submitted for publication.
\bibitem{sk-book} S. Kumano, 
   Nuclear Physics (KEK Physics Series, Vol.2, Kyoritsu Shuppan, 2015, in Japanese). 
\bibitem{qcd-collider-book} R. K. Ellis, W. J. Stirling, and B. R. Webber,
           QCD and Collider Physics (Cambridge University, 1996).
\bibitem{nuclear-summary} D. F. Geesaman, K. Saito, and A. W. Thomas,
           Ann. Rev. Nucl. Part. Sci. {\bf 45} (1995) 337;
   L. Frankfurt, V. Gusey, and M. Strikman, Phys. Rept. {\bf 512} (2012) 255.
\bibitem{some-npdfs} M. Hirai, S. Kumano, and T. -H. Nagai, 
                  Phys. Rev. C {\bf 76} (2007) 065207;
     K. J. Eskola, H. Paukkunen, and C. A. Salgado, 
                JHEP {\bf 04} (2009) 065;
     D. de Florian, R. Sassot, P. Zurita, and M. Stratmann, 
                Phys. Rev. D {\bf 85} (2012) 074028;
     K. Kovarik {\it et al.}, arXiv:1509.00792.
\bibitem{nCTEQ-nu} I. Schienbein {\it et al}.,
             Phys. Rev. D {\bf 77} (2008) 054013; D {\bf 80} (2009) 094004; 
     K. Kovarik {\it et al}., Phys. Rev. Lett. {\bf 106} (2011) 122301.
         The figure 5 was obtained by a personal communication with an author.
\bibitem{Kalantarians-2015} N. Kalantarians, talk at this
     10th International Workshop on Neutrino-Nucleus Interactions 
     in the Few-GeV Region, Osaka, Japan, Nov. 16-21, 2015.
\bibitem{Q2->0-theory}
  A. Donnachie and P. V. Landshoff, Z. Phys. C {\bf 61} (1994) 139;
  B. Z. Kopeliovich, Nucl. Phys. B (Proc. Suppl.) {\bf 139} (2005) 219;
  S. A. Kulagin and R. Petti, Phys. Rev. D {\bf 76} (2007) 094023.
\bibitem{by-2011} 
  A. Bodek and U.-K. Yang, arXiv:1011.6592.
\bibitem{fluka-2009} 
  G. Battistoni {\it et al.}, Acta Phys. Pol. B {\bf 40} (2009) 2431;
  see http://www.fluka.org for the FLUKA package.
\bibitem{summary-duality} W. Melnitchouk, R. Ent, and C. E. Keppel, 
            Phys. Rept. {\bf 406}  (2005) 127.
     The figure 6 was obtained by a personal communication with the authors.
\bibitem{lhc-npdfs} I. Helenius, H. Paukkunen, and K. J. Eskola,
            arXiv:1509.02798.
\bibitem{fermilab-e906} K. Nakano, talk at the 10th Circum-Pan-Pacific
      Spin Symposium on High Energy Spin Physics, 
      Taipei, Taiwan, Oct. 5-8, 2015.
\bibitem{ubar-dbar} S. Kumano, Phys. Rept. {\bf 303} (1998) 183;
                  G. T. Garvey and J.-C. Peng,
                       Prog. Part. Nucl. Phys. {\bf 47} (2001) 203;
	   J.-C. Peng and J.-W. Qiu, Prog. Part. Nucl. Phys. {\bf 76} (2014) 43.
\bibitem{minerva} J. Mousseau, J. G. Morfin, talks at this
     10th International Workshop on Neutrino-Nucleus Interactions 
     in the Few-GeV Region, Osaka, Japan, Nov. 16-21, 2015.
\end{thebibliography}
\end{document}